\documentclass[runningheads]{llncs}

\usepackage{graphicx}
\usepackage[inline]{enumitem} 
\usepackage{booktabs}
\usepackage{adjustbox}
\usepackage[hidelinks]{hyperref}

\begin{document}

\title{SmartReviews: Towards Human- and Machine-actionable Reviews}

\author{Allard Oelen\inst{1,2}\orcidID{0000-0001-9924-9153} \and
Markus Stocker\inst{2}\orcidID{0000-0001-5492-3212} \and
S\"oren Auer\inst{1,2}\orcidID{0000-0002-0698-2864}}

\authorrunning{Oelen et al.}

\institute{L3S Research Center, Leibniz University of Hannover, Hannover, Germany
\email{oelen@l3s.de}\and
TIB Leibniz Information Centre for Science and Technology, Hannover, Germany\\
\email{\{markus.stocker,soeren.auer\}@tib.eu}}

\newcounter{rtaskno}
\DeclareRobustCommand{\rtask}[1]{%
   \refstepcounter{rtaskno}%
   \textbf{R\thertaskno\label{#1}}}

\maketitle 

\begin{abstract}
Review articles summarize state-of-the-art work and provide a means to organize the growing number of scholarly publications. However, the current review method and publication mechanisms hinder the impact review articles can potentially have. Among other limitations, reviews only provide a snapshot of the current literature and are generally not readable by machines. In this work, we identify the weaknesses of the current review method. Afterwards, we present the \textit{SmartReview} approach addressing those weaknesses. The approach pushes towards semantic community-maintained review articles. At the core of our approach, knowledge graphs are employed to make articles more machine-actionable and maintainable. 

\end{abstract}

\section{Introduction}
\label{section:introduction}
\begin{figure*}[t]
    \centering
    \includegraphics[width=0.85\textwidth]{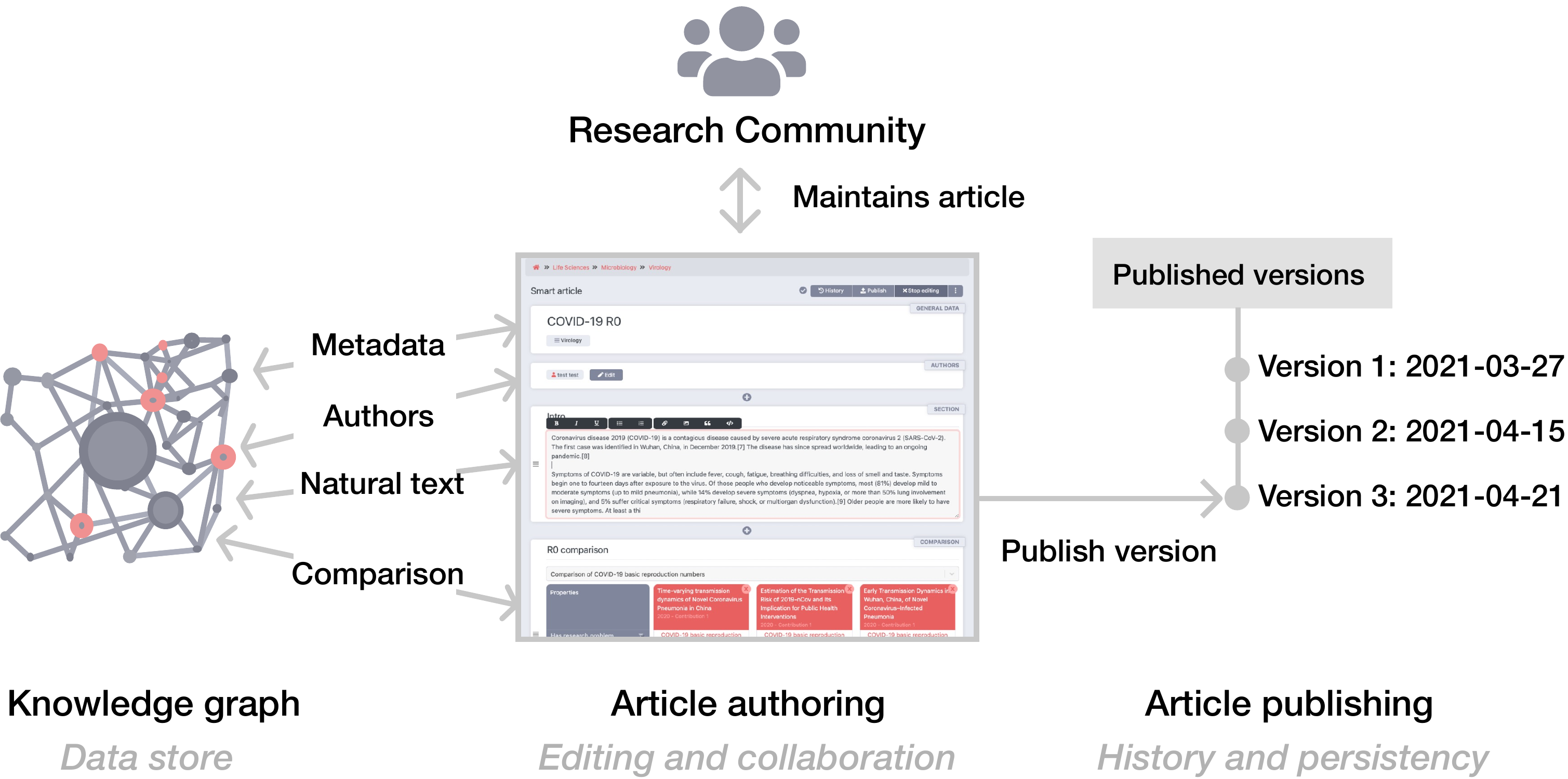}
    \caption{Overview of key concepts of the SmartReview approach. Review articles are built on top of a knowledge knowledge graph.
    }
    \label{fig:teaser}
\end{figure*}

The number of published scholarly articles remains to grow steadily~\cite{Jinha2010a}. Scholarly communication mainly relies on document-based methods, often using PDF files to communicate and share knowledge. This traditional document-based communication method has several limitations either caused by the PDF format itself or by the document-based approach in general~\cite{Kuhn2016DecentralizedNanopublications}. We distinguish research articles and review articles. The former presents original research contributions while the latter reviews contributions from other work~\cite{Wee2016}. Review articles, in particular, are severely limited in their scope and reach due to the static nature of publications. They give extensive overviews of research for a particular domain, but do so merely for a period of time up to when the review is conducted. Because of the static nature of published articles, updating reviews is either cumbersome or not possible at all~\cite{NEPOMUCENO201940,Wohlin2020}. This results in review articles being outdated soon after they are published, especially in research domains that face rapid (technology) evolution. In this work, we reimagine scholarly publishing for reviews by presenting a collaborative and community-maintained approach that aims to address the limitations and weaknesses of document-based communication. At the core of our approach, we leverage knowledge graphs for representing content of articles in a semantic machine-actionable manner. The approach and its main concepts are summarized in Fig.~\ref{fig:teaser}. A prototype of the approach is implemented in the Open Research Knowledge Graph (ORKG)~\cite{jaradeh2019open} and is available online\footnote{\url{https://www.orkg.org/orkg/smart-reviews}}. In summary, this work makes the following research contributions: 
\begin{enumerate*}[label=(\roman*)]
\item Analysis of the limitations and weaknesses of the current review method. 
\item Presentation of the SmartReview concept and approach to address the limitations.
\end{enumerate*}

\section{Weaknesses of Current Approach}
\label{section:related-work}

The current approach of authoring and publishing review articles has multiple weaknesses. The weaknesses are identified based on previous work, in particular from~\cite{Oelen2020a}. 

\textbf{Lacking updates.}
Once an article is published, it is generally not updated~\cite{MENDES2020110607}. This is caused either by lacking incentives from the author's perspective or due to technical limitations. For most research articles, this is acceptable. After all, if new results are available, it provides an opportunity to publish a new article building upon previous work. However, specifically for review articles this implies that the articles are outdated soon after they are published.

\textbf{Lacking collaboration.}
Reviews include research articles created by numerous authors. With the current review method, only the viewpoint of the review authors is considered and not from the community as a whole. This potentially imposes biases and hinders the objectiveness of the discussion of the reviewed work. Schmidt et al. found that a considerable amount of evaluated narrative review articles for the medical domain was severely biased~\cite{schmidt2005mites}. 

\textbf{Limited coverage.}
Authoring review articles is a resource intensive activity, which is generally more cumbersome than writing a research article~\cite{Wee2016}. Therefore, reviews are often only conducted for relatively popular domains and are lacking for less popular domains. Since review articles are an important factor for the development of research domains~\cite{Webster2002}, the lack of review articles can potentially hinder the evolution of a domain. 

\textbf{Lacking machine-actionability.}
The most frequently used format for publishing scholarly articles is PDF, which is hard to process for machines~\cite{Klampfl2014}. PDF files focus on visual presentation specifically designed for human consumption. Nowadays, machine consumption of PDF files relies on machine learning techniques and is often limited to parsing the article's metadata~\cite{Nasar2018,Lipinski2013}. 

\textbf{Limited accessibility.}
Documents published in PDF format are often inaccessible to readers with disabilities~\cite{Ahmetovic2018}. PDF documents focus on the visual representation of documents instead of a structured representation, which hinders accessibility~\cite{Darvishy2018}. 

\textbf{Lacking overarching systematic representation.}
Generally, there is no systematic representation of concepts used in articles, which means scholarly publishing does not use related web technologies to their full potential~\cite{Shotton2009a}. This has several implications and potentially causes redundancy and ambiguity across scholarly articles.

\section{The SmartReview Approach}
\label{section:approach}

Based on the identified weaknesses, we devise the SmartReview approach and determine its six definitorial dimensions. Each dimension presents system requirements that define how the dimension is addressed. Requirements are formulated using the FunctionalMASTeR template~\cite{TheSophists2016}. 

\noindent\textbf{Article Updates.}
It should be possible to update review articles once published, resulting in ``living" documents~\cite{Shanahan2015}. The individual versions should be citable and it should be clear which version of the article is cited. Additionally, readers should be able to see which parts of the articles have changed across versions. Based on these criteria, we formulate the following requirements:
\begin{itemize}[leftmargin=0.9cm]
  \item[\rtask{requirement:updates-updatable}] SmartReviews shall provide researchers the ability to update articles. 
  \item[\rtask{requirement:updates-persistency}] SmartReviews shall persist all versions of published articles.
  \item[\rtask{requirement:updates-diff-view}] SmartReviews shall provide researchers with the ability to compare different versions of the same article (i.e., diff view).
 \end{itemize}

\noindent\textbf{Collaboration.} To fully support community collaboration for review articles, they should be editable by anyone within the community. To ensure no work is getting lost (e.g., removed by another author), it should be possible to go back in time and compare different versions (as described in~R\ref{requirement:updates-diff-view}). 
\begin{itemize}[leftmargin=0.9cm]
  \item[\rtask{requirement:collaboration-open-to-all}]
  SmartReviews shall provide any researcher with the ability to contribute to articles. 
  \item[\rtask{requirement:collaboration-acknowledgements}]
  SmartReviews shall list all contributors in the acknowledgements. 
\end{itemize}

\noindent\textbf{Coverage.} To increase the review coverage for less popular domains, the entry barrier for creating and updating SmartReviews should be low (related to~R\ref{requirement:collaboration-open-to-all}). SmartReviews can be created even if only a limited amount of articles are reviewed. This is achieved by decoupling publishing (i.e., peer-reviewed publishing in a journal or conference) and authoring of articles.
\begin{itemize}[leftmargin=0.9cm]
  \item[\rtask{requirement:coverage-no-peer-review}]
  SmartReviews shall provide researchers with the ability to create articles without the need for an a priori peer review.
\end{itemize}

\noindent\textbf{Machine-actionability.} In order to improve machine-actionability, a systematic and structured representation in a knowledge graph should be used for knowledge representation. The resources defined within the knowledge graph serve as building blocks to create the article. This structured data is supplemented by natural text sections. To improve machine-actionability, natural text sections are complemented by types describing their contents. 
\begin{itemize}[leftmargin=0.9cm]
  \item[\rtask{requirement:actionability-graph}] SmartReviews shall use a knowledge graph as data source for articles. 
  \item[\rtask{requirement:actionability-natural-text}] SmartReviews shall semantically type and structure natural language text sections.
  \item[\rtask{requirement:actionability-formats}]
  SmartReviews shall provide machine-actionable formats (i.e., RDF, JSON-LD).
\end{itemize}

\noindent\textbf{Accessibility.} Most accessibility issues originate from the PDF format in which most articles are published. By publishing the articles in HTML instead, the article is already more accessible. Furthermore, by adhering to the Web Content Accessibility Guidelines (WCAG)~\cite{Caldwell2008}, the accessibility is further improved. 
\begin{itemize}[leftmargin=0.9cm]
  \item[\rtask{requirement:accesibility-html}]
  SmartReviews shall publish articles in HTML format.
  \item[\rtask{requirement:accesibility-wcag}]
  SmartReviews shall follow WCAG guidelines. 
\end{itemize}

\noindent\textbf{Systematic representation.} Review articles often use tabular representations for comparing research contributions from different articles. SmartReviews should focus on these comparison tables, and encourage researchers to use these tables to devise a structured description of the reviewed articles. 

\begin{itemize}[leftmargin=0.9cm]
  \item[\rtask{requirement:representation-comparison}] SmartReviews shall devise a structured comparison of reviewed work. 
  \item[\rtask{requirement:representation-resources}] SmartReviews shall support linking existing resources and properties from the knowledge graph. 
\end{itemize}

\section{Conclusion}
\label{section:conslusion}
The current review method suffers from numerous weaknesses which we described in this work. Based on the identified weaknesses, we devised the SmartReview approach. This approach proposes a collaborative community-maintained method for authoring review articles. It employs a knowledge graph to support machine-actionable articles. Future work will focus on evaluation and implementation of the SmartReview approach. 

\subsubsection*{Acknowledgements}
This work was co-funded by the European Research Council for the project ScienceGRAPH (Grant agreement ID: 819536) and the TIB Leibniz Information Centre for Science and Technology.

\newpage

\bibliographystyle{splncs04}
\bibliography{references-mendeley}

\end{document}